\documentclass[preprint,12pt,number]{elsarticle}

\biboptions{sort&compress}

\usepackage{amssymb}

\usepackage[colorlinks,urlcolor=blue,citecolor=blue,linkcolor=blue,pdfstartview=FitH]{hyperref}
\usepackage{gensymb}
\usepackage{array,multirow}
\usepackage{amsopn}

\journal{and published in Comput. Mater. Sci.}

\DeclareMathOperator*{\sumsum}{\sum\sum} 

\newcommand{\young}{E}
\newcommand{\conc}{n}
\newcommand{\stress}{\sigma_{zz}}
\newcommand{\strain}{\varepsilon_{zz}}
\newcommand{\tensilestrength}{\sigma_{\mathrm{TS}}}
\newcommand{\abbrev}[1]{{\footnotesize #1}}
\newcommand{\eqref}[1]{(\ref{#1})}

\newcommand{\BLYP}{{\footnotesize{B3LYP}}} 
\newcommand{\CASSCF}{{\footnotesize{CASSCF}}} 
\newcommand{\CCSDT}{{\footnotesize{CCSDT}}} 
\newcommand{\CI}{{\footnotesize{CI}}} 
\newcommand{\DFT}{{\footnotesize{DFT}}} 
\newcommand{\ECP}{{\footnotesize{ECP}}} 
\newcommand{\FLAPW}{{\footnotesize{FLAPW}}} 
\newcommand{\GGA}{{\footnotesize{GGA}}} 
\newcommand{\MBPPCA}{{\footnotesize{MBPP}}} 
\newcommand{\MRCPA}{{\footnotesize{MRCP4}}} 
\newcommand{\MRSDCIQ}{{\footnotesize{MRSDCI}}} 
\newcommand{\PAW}{{\footnotesize{PAW}}} 
\newcommand{\SDQMBPT}{{\footnotesize{MBPT4}}} 
\newcommand{\UHF}{{\footnotesize{UHF}}} 
\newcommand{\USPEX}{{\footnotesize{USPEX}}} 
\newcommand{\USPP}{{\footnotesize{USPP}}} 

\begin{document}
\date{January, 2016}

\begin{frontmatter}

\title{Interatomic Fe--H potential for irradiation and embrittlement simulations} 

\author[monash,helsinki]{Pekko Kuopanportti}\ead{pekko.kuopanportti@monash.edu}
\author[cea]{Erin Hayward}\ead{erin@gatech.edu}
\author[cea]{Chu-Chun Fu}\ead{chuchun.fu@cea.fr}
\author[helsinki]{Antti Kuronen}\ead{antti.kuronen@helsinki.fi}
\author[helsinki]{Kai Nordlund}\ead{kai.nordlund@helsinki.fi}

\address[monash]{School of Physics and Astronomy, Monash University, Victoria 3800, Australia}
\address[helsinki]{Department of Physics, University of Helsinki, P.O. Box 43, FI-00014 Helsinki, Finland}
\address[cea]{CEA, DEN, Service de Recherches de M\'{e}tallurgie Physique, F-91191 Gif-sur-Yvette, France}

\begin{abstract}
The behavior of hydrogen in iron and iron alloys is of interest in many fields of physics and materials science. To enable large-scale molecular dynamics simulations of systems with Fe--H interactions, we develop, based on density-functional theory calculations, an interatomic Fe--H potential in the Tersoff--Brenner formalism.  The obtained analytical potential is suitable for simulations of H in bulk Fe as well as for modeling small FeH molecules, and it can be directly combined with our previously constructed potential for the stainless steel Fe--Cr--C system. This will allow simulations of, e.g., hydrocarbon molecule chemistry on steel surfaces. In the current work, we apply the potential to simulating hydrogen-induced embrittlement in monocrystalline bulk Fe and in an Fe bicrystal with a grain boundary. In both cases, hydrogen is found to soften the material.
\end{abstract}


\end{frontmatter}

\section{Introduction}

Hydrogen, although not soluble in iron in equilibrium, can be introduced into it by irradiation, nuclear decay, or chemical processes. Hydrogen is well known to cause embrittlement in iron and steel~\cite{Ori1978.ARMS8.327,Kim1986.MSE77.75,Ber1970.MSE6.1,Zho2000.PRB62.13938,Nag2001.MetMatTransA32.339,Tat2003.PRB67.174105,Gen2005.MatTrans46.756,Tia2011.JPCM23.015501}, which is a serious issue in, e.g., the automotive and nuclear industries. In the former, the high mechanical resistance desired from the body steels must often be traded off against their increased susceptibility to hydrogen embrittlement~\cite{Ori1985_book,Lov2012.MetMatTransA43.4075,Koy2012.CorrosionScience54.1,Koy2012.CorrosionScience59.277}, while the nuclear processes in the latter will, on long time scales, induce hydrogen buildup in the reactor steels~\cite{Alt1965.JNM16.68,Hir1976.Corrosion32.3,Sco1994.JNM211.101,Gre1994.JNM216.29}. Moreover, the recent changes in the design of the ITER fusion reactor are to render some of its steel components directly exposed to the fusion plasma~\cite{CarbonoutfromITERICFRM}, making it important to study how the energetic H isotopes escaping from the plasma interact with Fe.

Atomic-level molecular dynamics~(MD) simulations have proven to be a good tool for examining irradiation effects~\cite{Ave98,Nor13a}, mechanical properties of materials~\cite{Keb97,Sch99,Bri05}, and plasma--wall interactions~\cite{Sal99,Nor13b}. The key physical input for MD is the interatomic potential.  Since steels by definition contain Fe and C~\cite{Callister}, simulations of H effects in steels require, at a minimum, a potential that can describe all interactions in the ternary Fe--C--H system.

In this work, we develop a potential for Fe--H interactions in the same reactive Tersoff--Brenner formalism~\cite{Ter1988.PRB37.6991,Bre1990.PRB42.9458,Alb01b} we used previously to construct a potential for the stainless steel Fe--Cr--C system~\cite{Hen10}. The potential is fitted to a database of properties of FeH molecules and H in bulk Fe, obtained from literature and our own density functional theory~(DFT) calculations.  By using already available C--H parameters~\cite{Bre1990.PRB42.9458,Jus05}, the potentials can be directly combined to model the entire ternary Fe--C--H system.  The potential allows simulating H in bulk Fe as well as ion irradiation and chemical reactivity of H at Fe and Fe--C surfaces. We demonstrate its use in simulating hydrogen-induced softening in bulk Fe and in Fe grain boundaries.

Fe--H potentials have been already devised using the embedded-atom method~\cite{Lee2007.AcMat55.6779,Ram2009.PRB79.174101}. However, its associated functional form cannot realistically describe the C--H bonding chemistry~\cite{Bre1990.PRB42.9458}. Since our aim is to obtain a potential for the entire Fe--C--H system, we choose to develop the Fe--H potential in the Tersoff--Brenner formalism which allows combining the Fe--H part with both Fe and C interactions, similar to what was done earlier for the W--C--H system~\cite{Jus05}.

The remainder of this article is organized as follows. In Section~\ref{sc:methods}, we summarize the Tersoff--Brenner potential formalism and describe our fitting procedure. Section~\ref{sc:obtained_potential} presents the obtained Fe--H potential and evaluates its performance against experimental and \emph{ab initio} data. In Section~\ref{sc:tensile_test}, we employ the potential in tensile-test simulations of hydrogen-containing iron. We discuss the implications and limitations of the study in Section~\ref{sc:discussion}. Finally, Section~\ref{sc:summary} concludes the article with a brief summary.

\section{Potential formalism and fitting procedure}\label{sc:methods}

The reactive Tersoff--Brenner formalism~\cite{Ter1988.PRB37.6991,Bre1990.PRB42.9458,Alb01b} used in this work originates from the concept of bond order proposed by Pauling~\cite{Pau1960.book}, and it has been shown~\cite{Alb2002.PRB65.195124} to resemble both the tight-binding scheme~\cite{Cle93.PRB48.22} and the embedded-atom method~\cite{Daw1984.PRB29.6443,Bre1989.PRL63.1022}. Since the formalism has been described extensively elsewhere~\cite{Alb2002.PRB65.195124,Alb2002.PRB66.035205,Nor2003.JPCM15.5649}, we will give here only a brief overview.

The total cohesive energy $E_\mathrm{c}$ of the system is written as a sum of individual bond energies: 
\begin{equation}\label{eq:ene}
E_\mathrm{c}=\sumsum_{i < j} f^\mathrm{c}_{ij} \left(r_{ij}\right)\left[V_{ij}^\mathrm{R}\left(r_{ij}\right) - \frac{b_{ij}+b_{ji}}{2} V_{ij}^\mathrm{A}\left(r_{ij}\right) \right],
\end{equation} 
where $r_{ij}$ is the distance between atoms $i$ and $j$, $f^\mathrm{c}$ is a cutoff function for the pair interaction, $V^\mathrm{R}$ is a repulsive and $V^\mathrm{A}$ an attractive pair potential, and $b_{ij}$ is a bond-order term that describes three-body interactions and angularity. The pair potentials are of the Morse-like form
\begin{eqnarray}
\label{eq:pair_potentials}
V^\mathrm{R}_{ij}\left(r\right) &=&\frac{D_{0,ij}}{S_{ij}-1} \exp\left[- \sqrt{2S_{ij}}\beta_{ij}
\left(r-r_{0,ij}\right)\right], \\ 
V^\mathrm{A}_{ij}\left(r\right) &=&\frac{S_{ij}D_{0,ij}}{S_{ij}-1} \exp\left[-\frac{\sqrt{2}\beta_{ij}}{\sqrt{S_{ij}}}
\left(r-r_{0,ij}\right)\right],
\end{eqnarray} 
where $D_0$ and $r_0$ are the bond energy and length of the dimer molecule, respectively. The parameter $\beta$ is related to the ground-state vibrational frequency $\omega$ and the reduced mass $\mu$ of the dimer according to 
\begin{equation} 
\beta_{ij}=\frac{\sqrt{2\mu_{ij}}\pi\omega_{ij}}{\sqrt{D_{0,ij}}}.
\end{equation}
The bond-order term is given by 
\begin{equation}
b_{ij} = \frac{1}{\sqrt{1+\chi_{ij}}},
\end{equation}
where 
\begin{equation}
\chi_{ij}=\sum_{k\left(\neq i,j\right)} f^\mathrm{c}_{ij}\left(r_{ij}\right) g_{ik}\left(\theta_{ijk}\right)\exp\left[\alpha_{ijk}\left(r_{ij}-r_{ik} \right)\right].
\end{equation} 
Here $\theta_{ijk}$ is the angle between the vectors $\mathbf{r}_{ij}= \mathbf{r}_{j} - \mathbf{r}_{i}$ and $\mathbf{r}_{ik}$, and the angular function is defined as 
\begin{equation}\label{eq:g}
g_{ik}\left(\theta_{ijk}\right)=\gamma_{ik}\left[1+\frac{c_{ik}^2}{d_{ik}^2}-\frac{c_{ik}^2}{d_{ik}^2+\left(h_{ik}+\cos\theta_{ijk}\right)^2}\right],
\end{equation}
where $\gamma$, $c$, $d$, and $h$ are adjustable parameters. The range of the interaction is restricted to the next-neighbor sphere by the cutoff function 
\begin{equation}\label{eq:cutoff} 
f^\mathrm{c}_{ij}\left(r\right) = \left\{ \begin{array}{l r} 1, & r\leq R_{ij}-D_{ij}, \\ \frac{1}{2}-\frac{1}{2}\sin\frac{\pi\left(r-R_{ij}\right)}{2D_{ij}}, & |r-R_{ij}|\leq D_{ij}, \\ 0, & r\geq R_{ij}+D_{ij}, \end{array} \right.
\end{equation} 
where $R$ and $D$ determine the locus and width of the cutoff interval.

If the analytical potential is used for modeling nonequilibrium phenomena involving short-distance interactions, such as high-energy particle irradiation processes or melting, the short-range part of the potential must be adjusted to include a strong repulsive core that follows, i.a., from the Coulomb repulsion between the positively charged nuclei. To this end, the potential is modified in the manner already used for other Tersoff-like many-body
potentials~\cite{Alb2002.PRB65.195124,Nor1996.PRL77.699}: The total potential $V_\mathrm{tot}$ is constructed by joining the universal Ziegler--Biersack--Littmark potential $V_\mathrm{ZBL}$~\cite{Zie1985_book} with the equilibirium potential $V_\mathrm{eq}$ using
\begin{equation}\label{eq:total_potential}
V_\mathrm{tot}\left(r\right)=V_\mathrm{ZBL}\left(r\right)\left[1-F\left(r\right)\right]+V_\mathrm{eq}\left(r\right)F\left(r\right),
\end{equation}
where $V_\mathrm{eq}$ is the potential implied by Eq.~\eqref{eq:ene} and $F$ is the Fermi function $F\left(r\right)=\left\{1+\exp\left[-b_\mathrm{F}\left(r-r_\mathrm{F}\right)\right]\right\}^{-1}$. The values of the
parameters $b_\mathrm{F}$ and $r_\mathrm{F}$ are chosen manually such that the potential is essentially unmodified at equilibrium and longer bonding distances and that a smooth fit at short separations with no spurious minima is obtained for all realistic coordination numbers.

In order to devise a well-performing Fe--H potential in the Tersoff--Brenner formalism, we use the following fitting procedure: The parameter sets for the H--H and Fe--Fe interactions are taken unchanged from Refs.~\cite{Bre1990.PRB42.9458} and~\cite{Mul2007.JPCM19.326220}, respectively, so that only the parameter set for the Fe--H interactions is fitted. From the outset, we fix the parameters pertaining to the properties of the dimer FeH---i.e., $D_0$, $r_0$, and $\beta$---according to their experimentally observed values. To avoid unwanted side effects, we set the three-index parameters $\alpha_{ijk}$ to zero. The values of the remaining seven parameters ($S$, $\gamma$, $c$, $d$, $h$, $R$, and $D$) are then fitted to a structural database comprising the molecules FeH${}_2$ and FeH${}_3$, the stoichiometric FeH with the rock-salt crystal structure, and the energies of the lowest-lying hydrogen point defects in body-centered cubic (bcc) iron. The fitting is formulated as a nonlinear least-squares minimization problem, which we solve using the trust-region-reflective algorithm~\cite{Mor1983.SJSSC3.553,Byr1988.MaPr40.247,Bra1999.SJSC21.1} implemented in {\footnotesize MATLAB}~\cite{matlab}.

\section{Obtained potential}\label{sc:obtained_potential}

The optimized parameter values for the analytical Fe--H potential are given in Table~\ref{table:potential_parameters}. We also show the parameter values used for the H--H and Fe--Fe potentials; it should be noted, however, that during the fitting process, H--H and Fe--Fe interactions play a role only in the evaluation of the hydrogen point-defect energies.

\begin{table}
\caption{\label{table:potential_parameters}Tersoff--Brenner parameters~[Eqs.~\eqref{eq:ene}--\eqref{eq:total_potential}] for the Fe--H system. The H--H potential is taken from Ref.~\cite{Bre1990.PRB42.9458} and the Fe--Fe potential from
Ref.~\cite{Mul2007.JPCM19.326220}; the Fe--H potential is derived in this work. The parameters $\alpha_{ijk}$ are zero in all cases. In~Section~\ref{sc:tensile_test}, we use $R=3.5$\,\AA\ for the Fe--Fe interactions instead of $3.15$\,\AA.}
\begin{tabular}{l l r r r}
\\
& & \multicolumn{1}{c}{H--H} & \multicolumn{1}{c}{Fe--Fe} & \multicolumn{1}{c}{Fe--H} \\ \hline \vspace{-8pt} \\
$D_0$ &(eV) & 4.7509 & 1.5\hphantom{0000} & 1.630\hphantom{00} \\
$r_0$ &(\AA) & 0.7414 & 2.29\hphantom{000} & 1.589\hphantom{00} \\
$\beta$ &(\AA${}^{-1}$) & 1.9436 & 1.4\hphantom{0000} & 1.875\hphantom{00}\\
$S$ & & 2.3432 & 2.0693\hphantom{0} & 4.000\hphantom{00}\\
$\gamma$ & & 12.33\hphantom{00} & 0.01158 & 0.01332 \\
$c$ & & 0.0\hphantom{000} & 1.2899\hphantom{0} & 424.5\hphantom{0000} \\
$d$ & & 1.0\hphantom{000} & 0.3413\hphantom{0} & 7.282\hphantom{00} \\
$h$ & & 1.0\hphantom{000} &$-0.26$\hphantom{000} & $-0.1091$\hphantom{0} \\
$R$ &(\AA) & 1.40\hphantom{00} & 3.15\hphantom{000} & 2.497\hphantom{00} \\
$D$ &(\AA) & 0.30\hphantom{00} & 0.2\hphantom{0000} & 0.1996\hphantom{0} \\
$b_\mathrm{F}$ &(\AA${}^{-1}$) & 15.0\hphantom{000} & 2.9\hphantom{0000} & 16.0\hphantom{0000}\\
$r_\mathrm{F}$ &(\AA) & 0.35\hphantom{00} & 0.95\hphantom{000} & 1.0\hphantom{0000} \\
\hline
\end{tabular}
\end{table}

Table~\ref{table:properties} presents the fitting database together with the results from the analytical potential. The potential exactly reproduces the experimentally observed dimer properties that are also in good agreement with all the \emph{ab initio} calculations. For the linear trimer FeH${}_2$, the experimentally measured bond lengths in Refs.~\cite{Kor1999.JCP110.3861} and~\cite{Kor1996.JCP104.4859} are, respectively, 1.1\% and 2.1\% greater than the analytical prediction, while differing from each other by roughly the same relative amount. The analytical potential yields a bond length for the trigonal planar FeH${}_3$ molecule that falls between the values given by the different \emph{ab initio} calculations. The lattice constant $a$, bulk modulus $B$, and its pressure derivative $B'$ for the rock-salt FeH are also in line with the DFT results.

\begin{table*}
\caption{\label{table:properties} Properties of the Fe--H molecular and rock-salt phases as obtained from experiments, \emph{ab initio} calculations, and the analytical potential~(AP) derived in this work. The notation is as follows: $r_\mathrm{b}$, bond length; $k$, wave number for the ground-state vibrational frequency; $E_\mathrm{c}/N$, cohesive energy per atom; $a$, lattice constant; $B$, bulk modulus; $B'$, pressure derivative of the bulk modulus. For the abbreviations of the \emph{ab initio} methods, see Table~\ref{table:acronyms} in the appendix.}
\begin{tabular}{llrrrrr}
\\
& & \multicolumn{1}{c}{Expt.} & \multicolumn{3}{c}{\emph{Ab initio} calculations} & {AP}\\
 \cline{4-6} \vspace{-5.0 pt} \\
\multicolumn{2}{l}{FeH} &  & \CI\abbrev{/}\ECP\,\cite{Sod1990.JCP92.2478} & \MRCPA\,\cite{Tan2001.JCP115.4558} & \MRSDCIQ\,\cite{Tan2001.JCP115.4558} & \\
\hline \vspace{-12pt} \\ 
$r_\mathrm{b}$ &(\AA) &1.589$^{\rm a}$ &1.578 &1.596 &1.582 &1.589 \\
$E_\mathrm{c}/N$ &(eV) & $-0.815^{\rm b}$ &$-0.71$ & $-0.90$ & $-0.94$ & $-0.815$ \\
\vspace{8.0 pt} $k$ &($\mathrm{cm}^{-1}$) &1774$^{\rm c}$ &1701 & 1735 & 1778 & 1774\\
\multicolumn{2}{l}{FeH${}_2$ linear} & & \CASSCF\,\cite{Sie1984.JCP81.1373} & \CI\,\cite{Sie1984.JCP81.1373} & \BLYP\,\cite{Wan2009.JPCA113.551} & \\
\hline  \vspace{-12pt} \\ 
$r_\mathrm{b}$ &(\AA) & 1.648$^{\rm d}$ & 1.746 & 1.689 & 1.645 & 1.630 \\
\vspace{8.0 pt} $E_\mathrm{c}/N$ &(eV) &---\hphantom{$^{\rm a}$} &--- &--- &--- &$-0.875$ \\
\multicolumn{2}{l}{FeH${}_3$ planar} & & \UHF\,\cite{Bal2000.JPCA104.1597} & \SDQMBPT\,\cite{Bal2000.JPCA104.1597} & \CCSDT\,\cite{Bal2000.JPCA104.1597} & \\
\hline  \vspace{-12pt} \\ 
$r_\mathrm{b}$ &(\AA) &---\hphantom{$^{\rm a}$} & 1.667 & 1.603 & 1.609 & 1.619 \\
\vspace{8.0 pt} $E_\mathrm{c}/N$ &(eV) &---\hphantom{$^{\rm a}$} &--- &--- &--- & $-1.044$  \\ 
\multicolumn{2}{l}{FeH rock salt} & & \USPEX\,\cite{Baz2012.PhysU55.489} & \MBPPCA\,\cite{Els1998.JPCM10.5081} & \FLAPW\,\cite{Els1998.JPCM10.5081} & \\
\hline  \vspace{-12pt} \\ 
$a$ &(\AA) &---\hphantom{$^{\rm a}$} & 1.828 & 1.839 & 1.833 & 1.839 \\
$E_\mathrm{c}/N$ &(eV) &---\hphantom{$^{\rm a}$} &--- &--- &--- & $-3.518$ \\
$B$ &(GPa) &---\hphantom{$^{\rm a}$} & 270.8& 216 & 200 & 238.9 \\
$B'$ & &---\hphantom{$^{\rm a}$} &4.25 &3.7 &3.7 &4.749 \\
\hline
\\
\end{tabular} 
\\
\noindent $^{\rm a}$ Balfour \emph{et al.}~\cite{Bal1983.PS28.551}\\ 
\noindent $^{\rm b}$ Schultz and Armentrout~\cite{Sch1991.JCP94.2262}\\
\noindent $^{\rm c}$ Stevens \emph{et al.}~\cite{Ste1983.JCP78.5420}\\
\noindent $^{\rm d}$ K{\"o}rsgen \emph{et al.}~\cite{Kor1999.JCP110.3861} 
\end{table*}

Table~\ref{table:defects} lists the formation energies of hydrogen point defects as obtained using DFT and our analytical potential. The formation energies are defined as
\begin{equation}
E_\mathrm{f} = E_\mathrm{def}\left(N_\mathrm{Fe},N_\mathrm{H}\right) - N_\mathrm{Fe} E_\mathrm{c}\left(\mathrm{Fe}\right) - N_\mathrm{H} E_\mathrm{c}\left(\mathrm{H}\right),
\end{equation}
where $E_\mathrm{def}\left(N_\mathrm{Fe},N_\mathrm{H}\right)$ is the total cohesive energy of the defect-containing cell with $N_\mathrm{Fe}$ iron and $N_\mathrm{H}$ hydrogen atoms; $E_\mathrm{c}\left(\mathrm{Fe}\right)$ and $E_\mathrm{c}\left(\mathrm{H}\right)$ are the atomic cohesive energies of bcc iron and the H${}_2$ molecule, respectively. For the analytical potentials of Table~\ref{table:potential_parameters}, $E_\mathrm{c}\left(\mathrm{Fe}\right)= -4.280\,\textrm{eV}$ and $E_\mathrm{c}\left(\mathrm{H}\right)=-2.375\,\textrm{eV}$. A summary of the DFT methods is provided in the Appendix.

\begin{table*}
\caption{\label{table:defects} Experimental and theoretical formation energies $E_\mathrm{f}$ in units of eV for the most relevant hydrogen point defects in bcc iron: the tetrahedral (T) and octahedral (O) interstitials and the substitutional defect. Theoretical values are given for both unrelaxed and relaxed atomic configurations. The penultimate column lists our DFT results, and the last column shows the values calculated from our analytical potential~(AP). For the DFT methods, see the appendix.}
\begin{tabular}{lrrrrr}
\\
&  & \multicolumn{3}{c}{DFT calculations} & \\
 \cline{3-5}
\vspace{-7.0pt} \\
Defect & Expt.\,\cite{Hir1980.MetallTransA11A.861} & {{\PAW}\,\cite{Jia2004.PRB70.064102}} &
{{\USPP}\,\cite{Miw2002.PRB65.155114}} & {This w.} & {AP}  \\
 \hline
Unrelaxed T interst.&--- &0.29 &--- &0.484 &0.515 \\
Relaxed T interst.&0.296 &0.20 &0.30 &0.234 &0.240 \\
Unrelaxed O interst.&--- &0.76 &--- &0.822 &1.186 \\
Relaxed O interst.&--- &0.33 &--- &0.259 &0.256 \\
Unrelaxed substit.&--- &--- &--- &2.855 &4.027 \\
Relaxed substit.&--- &--- &--- &2.526 &3.145 \\ 
\hline
\end{tabular}
\end{table*}

Due to the low solubility and high mobility of hydrogen in iron, as well as the high probability of trapping at defect sites at low temperatures, little direct evidence for the site occupancy exists. Indirect evidence indicates that H resides in the tetrahedral (T) site of bcc Fe, with an experimental value of $0.296$\,eV per atom for the dissolution energy of H in Fe~\cite{Hir1980.MetallTransA11A.861}. According to the DFT results in Table~\ref{table:defects}, the T site is more stable, both for the unrelaxed and relaxed structures. The DFT calculations also indicate that the octahedral (O) site occupancy gains significantly more stabilization from lattice distortion than the T site does. This is because the O site undergoes a greater structural distortion than the T site, which can be understood heuristically by considering the sizes of the two sites: Using the lattice constant $a_\mathrm{Fe}=2.86$\,\AA, the radii of the T and O sites are 0.36\,\AA\ and 0.19\,\AA, respectively. The hydrogen atom has a covalent radius of 0.37\,\AA, so it fits better in the T site and causes smaller lattice distortions than in the O site. The same argument also explains why the energy of the substitutional defect decreases only slightly when relaxed. The analytical potential qualitatively reproduces this behavior and yields very good quantitative agreement for all three \emph{relaxed} defect energies.

Regarding the diffusion of hydrogen in bcc iron, Jiang and Carter~\cite{Jia2004.PRB70.064102} used DFT to obtain the Arrhenius equation for the diffusion coefficient, $D_\mathrm{diff}=D_\mathrm{diff}^{(0)} \exp\left(-E_\mathrm{a}/k_\mathrm{B} T\right)$, where $D_\mathrm{diff}^{(0)}=1.5\times 10^{-7}\,\textrm{m}{}^2\textrm{s}{}^{-1}$ and the zero-point-energy-corrected activation energy $E_\mathrm{a} = 0.042\,\textrm{eV}$ corresponds to direct hopping between two neighboring T sites. Our DFT calculations yield $E_\mathrm{a} = 0.044\,\textrm{eV}$ for the same transition path~\cite{Hay2013.PRB87.174103}. On the other hand, since H is easily trapped by impurities in Fe, the diffusion coefficients of H in Fe from laboratory measurements show a large scatter: Hayashi and Shu~\cite{Hay2000.SSP73-75.65} compile experimental values of $E_\mathrm{a}$ in the range from $0.035$\,eV to $0.142$\,eV.  Using our analytical potential and the nudged elastic band method~\cite{Mil1995.SurfaceScience324.305}, we get $E_\mathrm{a}=0.112\,\textrm{eV}$ for the nearest-neighbor $\mathrm{T}\to \mathrm{T}$ path. This value and the two DFT results all fall within the experimental range. 

\section{Effect of hydrogen on tensile testing of iron}\label{sc:tensile_test}

\begin{figure}
\begin{center}
\includegraphics[ width=270pt, keepaspectratio]{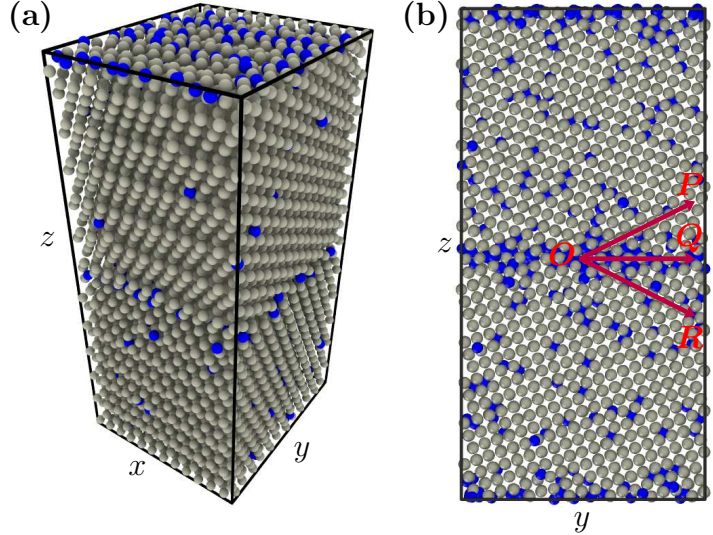}
\end{center}
\caption{\label{fig:cell} (a)~Perspective and (b)~side views of the computational cell used for the grain-boundary system with an atomic hydrogen concentration of 2.0\%. Iron atoms ($N_\mathrm{Fe}=8640$) are shown in gray (light) and hydrogen atoms ($N_\mathrm{H}=260$) in blue (dark). The dimensions of the cell are $34\,\textrm{\AA} \times 38\,\textrm{\AA}\times 78\,\textrm{\AA}$, and periodic boundary conditions are imposed in all three directions. The cell contains symmetric tilt boundaries in its top/bottom and middle sections. (b)~Neighboring grains are tilted about the $z$ axis by the angle $\angle POR = 53.13\degree$ with respect to each other; the vectors $\vec{OP}$ and $\vec{OR}$ correspond to equivalent lattice directions in the adjacent grains. The vector $\vec{OQ}$ lies along the grain boundary.}
\end{figure}

As a demonstration of possible applications of the derived Fe--H potential~(Table~\ref{table:potential_parameters}), we employ it in MD simulations~\cite{parcas} to investigate the effect of hydrogen impurity atoms on the stress--strain response of crystalline iron subjected to uniaxial tensile stress. We consider two types of computational cells of $N_\mathrm{Fe}=8640$ iron atoms, one consisting of a regular bcc lattice of $12\times 12\times 30$ unit cells and one containing a bcc bicrystal with a grain-boundary plane $\left(001\right)$ at its center; the latter is illustrated~\cite{ovito} in Fig.~\ref{fig:cell}. Both cells have $a_\mathrm{Fe}=2.86$\,\AA, and periodic boundary conditions are imposed in all three directions. The axis and angle of rotation for the grain boundary are chosen as $\left[100\right]$ and $53.13\degree$, so that  due to the periodic boundary conditions, the structure corresponds to a stack of symmetric tilt boundaries with a separation distance of $\sim\!\!39$\,\AA\ and a grain-boundary energy of $\left[E_\mathrm{cell} - N_\mathrm{Fe} E_\mathrm{c}\left(\mathrm{Fe}\right)\right]/2 A= 6.49\,\textrm{Jm}{}^{-2}$, where $E_\mathrm{cell}$ is the total energy of the computational cell and $A=286$\,\AA${} ^2$ is its cross-sectional area perpendicular to the $\left[001\right]$ direction.  In crystallographic notation~\cite{Lej2010.book}, the grain boundary can be described as  $53.13\degree \left[100\right]\left(0{\bar{1}}{\bar{2}}\right)/\left(0{\bar{1}}2\right)$.

To introduce the impurities into system, we randomly place a variable number $N_\mathrm{H}$ of hydrogen atoms into the computational cell, subject to the condition that the added atoms are at a minimum distance of 1.55\,\AA\ from the already existing atoms. For $N_\mathrm{H}$, we use the values 87, 174, 260, 346, 432, 519, 605, 691, 778, and 864, which corresponds to atomic hydrogen concentrations [defined as $\conc=N_\mathrm{H}/\left(N_\mathrm{Fe}+N_\mathrm{H}\right)$] of 0.0\%, 1.0\%, 2.0\%, 2.9\%, 3.9\%, 4.8\%, 5.7\%, 6.5\%, 7.4\%, 8.3\%, and 9.1\%, respectively.
We let the stresses in the hydrogen-containing cell relax to zero by evolving the system for 20\,ps at 300\,K while applying the Berendsen pressure control~\cite{Ber1984.JCP81.3684} in all directions. 

Next, exertion of uniaxial tension in the $z$ direction is modeled in a quasistatic, stepwise manner:  First, we increase the length $L_z$ of the simulation box by 0.02\,\AA, scaling the $z$ coordinates of all atoms by the ratio of the new and previous $L_z$. Second, we evolve the system for 50\,ps with fixed  $L_z$, while applying the Berendsen pressure control in the $x$ and $y$ directions and the Berendsen temperature control~\cite{Ber1984.JCP81.3684}  at 300\,K, and extract the axial normal stress $\stress$ as a time average over the last 25\,ps. These two steps are repeated 500 times, resulting in a maximum strain of 11--13\%. We carry out the whole procedure for different values of $N_\mathrm{H}$ and average the results for each $N_\mathrm{H}$ over ten independent initial configurations of H atoms.

Since the tensile-test simulations are carried at 300\,K, we have increased the value of the cutoff parameter $R$ for the Fe--Fe potential~\cite{Mul2007.JPCM19.326220} to 3.5\,\AA\ from the original 3.15\,\AA. Otherwise, the second-nearest neighbors of the bcc iron would---due to thermal vibrations---experience the onset of the cutoff function [Eq.~\eqref{eq:cutoff}], resulting in an unphysical increase in the Young's modulus of elasticity $\young = \stress / \strain$, where $\strain$ denotes the normal tensile strain. By extending $R$ to the middle of the second- and third-nearest-neighbor distances, this effect is avoided.

\begin{figure}
\begin{center}
\includegraphics[ width=250pt, keepaspectratio]{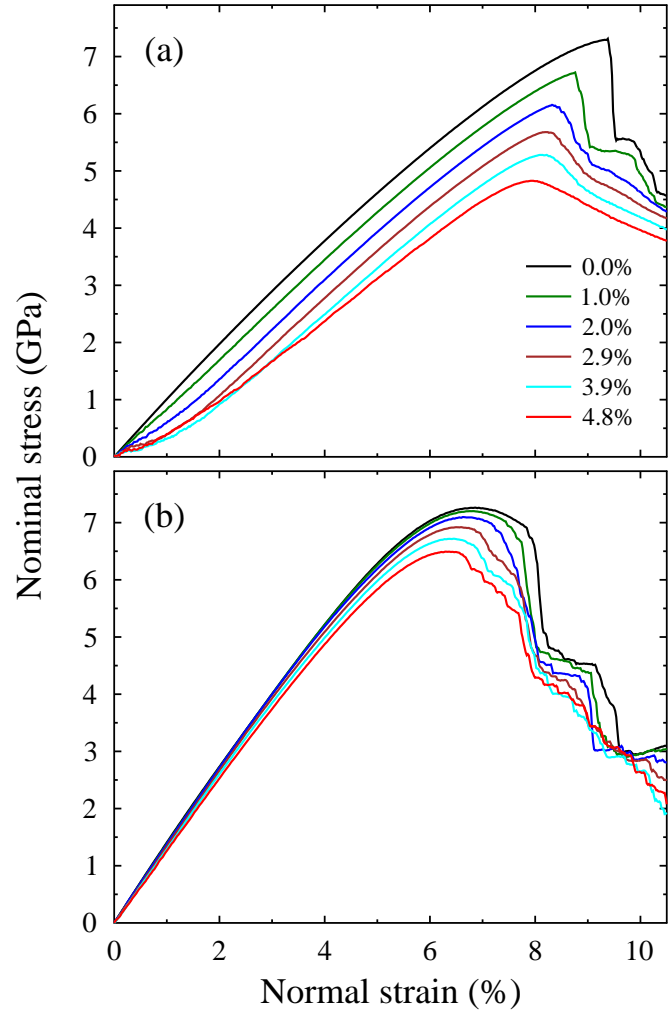}
\end{center}
\caption{\label{fig:stress--strain_bulk_gb} Uniaxial nominal stress $\stress$ as a function of normal tensile strain $\strain$ for (a)~monocrystalline bulk iron and (b)~an iron bicrystal with a symmetric tilt boundary~(Fig.~\ref{fig:cell}). The legend shows the atomic hydrogen concentration for each curve. Uniaxial tension is exerted along the $z$ axis, perpendicular to the grain-boundary plane.}
\end{figure}

\begin{figure}
\begin{center}
\includegraphics[ width=250pt, keepaspectratio]{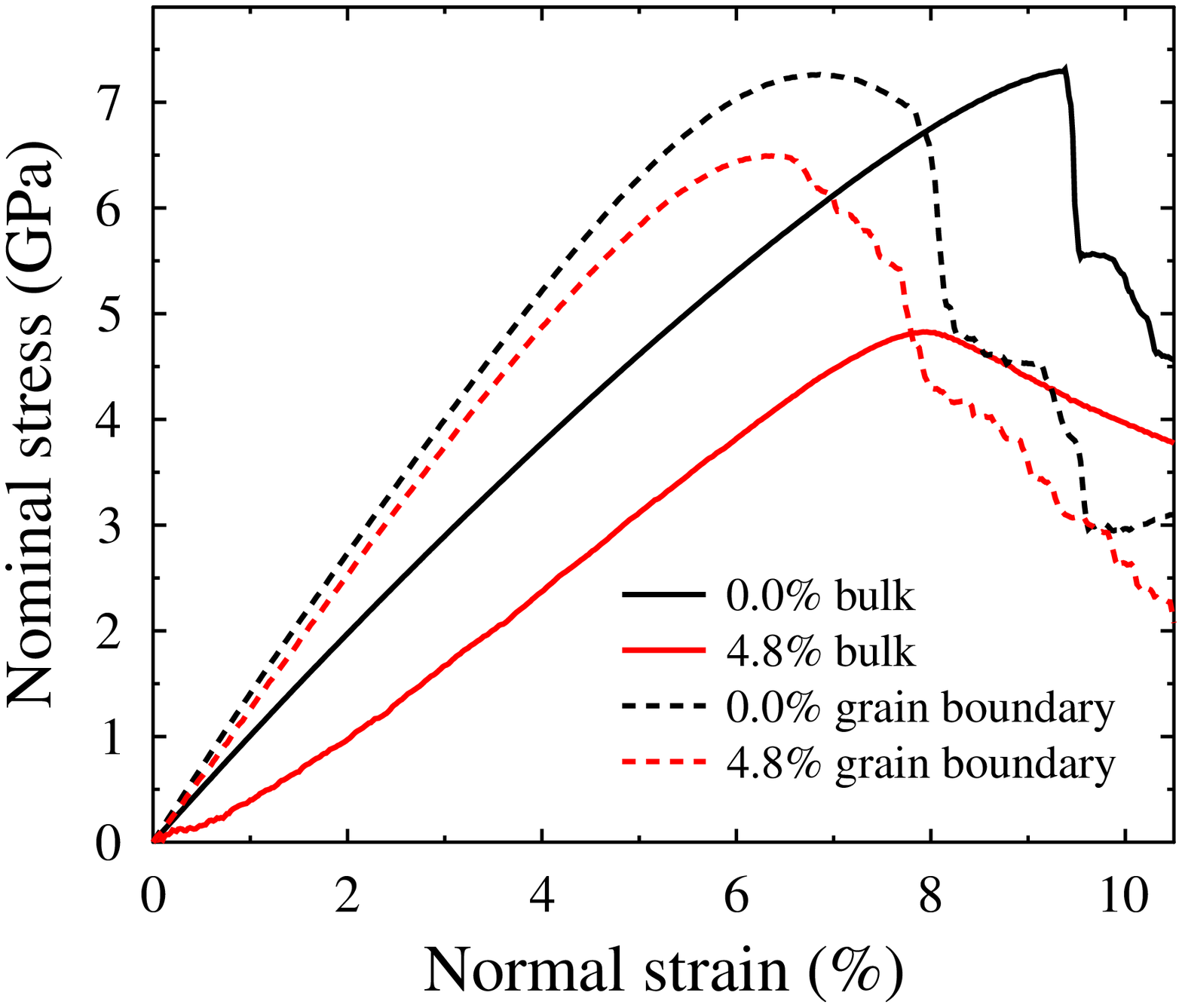}
\end{center}
\caption{\label{fig:stress--strain_both} Comparison of the uniaxial stress--strain responses of bulk~[see Fig.~\ref{fig:stress--strain_bulk_gb}(a)] and grain-boundary~[Fig.~\ref{fig:stress--strain_bulk_gb}(b)] iron crystals for atomic hydrogen concentrations of 0\% and 4.8\%.}
\end{figure}

Figure~\ref{fig:stress--strain_bulk_gb}(a) shows the average stress--strain curves for the regular monocrystalline (bulk) bcc iron, for six different atomic hydrogen concentrations. Figure~\ref{fig:stress--strain_bulk_gb}(b) depicts the corresponding curves for the grain-boundary system, and Fig.~\ref{fig:stress--strain_both} combines data from the two configurations.

To quantify the stress--strain response in the linear, elastic regime, we determine the Young's modulus $\young$ at different values of the hydrogen concentration $\conc$. This is done by performing a linear least-squares fit to each of the stress--strain curves in Fig.~\ref{fig:stress--strain_bulk_gb}. In the case of the bulk system, we  use the fitting intervals $\strain\in\left[0,0.010\right]$ for $\conc <2\%$, $\left[0.010,0.020\right]$ for $1\%<\conc <3\%$, and $\left[0.015,0.025\right]$ for $\conc >3\%$;  for the grain-boundary system, they are $\strain\in\left[0,0.010\right]$ for  $\conc <6\%$ and $\left[0.005,0.015\right]$ for $\conc >6\%$. The reason for not always starting the fitting interval from zero strain is that due to the high mobility of hydrogen in iron, there was some reorganization of the hydrogen atoms during the first few steps of the stretching procedure, producing nonlinear stress-strain behavior. As can be seen from Fig.~\ref{fig:stress--strain_bulk_gb}(a), the nonlinearity is particularly pronounced for bulk samples with large $\conc$.

\begin{figure}
\begin{center}
\includegraphics[ width=250pt, keepaspectratio]{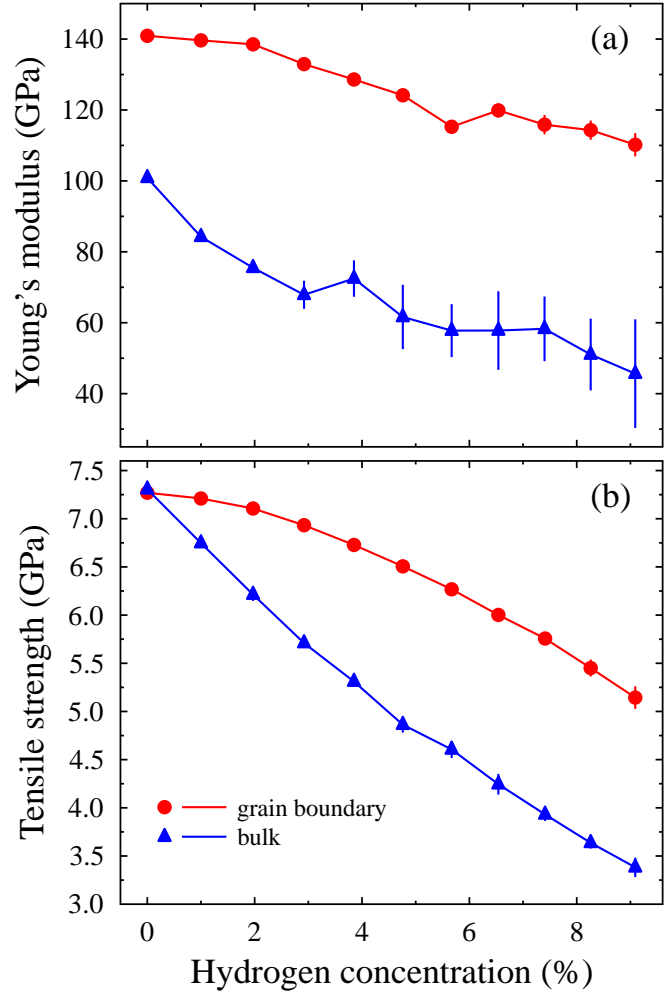}
\end{center}
\caption{\label{fig:modulus_and_yield_stress} (a)~Young's modulus $\young=\stress/\strain$ and (b)~the tensile strength $\tensilestrength$ as functions of the atomic hydrogen concentration for bulk bcc iron (triangles) and for the grain-boundary system (circles). The error bars are calculated as the standard deviation of ten simulations. The lines are guides to the eye.}
\end{figure}

Figure~\ref{fig:modulus_and_yield_stress}(a) shows the obtained Young's moduli at 300\,K for $\conc\in\left[0,0.1\right]$~\cite{note_extra_data}. The error bars are calculated as the standard deviation of each set of ten simulations. Considering first pure iron at  0\,K, its Young's moduli can be determined directly from the elastic moduli predicted by the potential~\cite{Mul2007.JPCM19.326220,Nye.book}. This gives the values 115\,GPa for the bulk bcc and 164\,GPa for a bcc system in which the lattice is rotated by the same angle and in the same direction as in our grain-boundary system. When the temperature is increased, the Young's moduli are expected to decrease. For pure iron at 300\,K, our simulations yield 101\,GPa for the bulk and 141\,GPa for the grain-boundary system. The experimental value~\cite{Adams2006} for the bulk system at 300\,K is 132\,GPa~\cite{note_discrepancy}. 

For both configurations, hydrogen is observed to induce softening of the material, i.e., to reduce $\young$. The effect is noticeably stronger for the bulk system: when the hydrogen concentration increases from zero to  9.1\%, the Young's modulus of the bulk system decreases by 55\%, whereas for the grain-boundary system the decrease is only  22\%. One possible explanation for this difference is that most of the hydrogen atoms in the grain-boundary system resided within $5$\,\AA\ from one of the boundaries. Therefore, the concentration of hydrogen in the \emph{intact} lattice was significantly lower than the nominal hydrogen concentration $\conc$ (e.g., at $\conc=0.091$ it was less than 4\%), while in the bulk system, these two quantities were obviously equal and the H atoms were homogeneously distributed throughout the computational cell.

Let us next investigate the extreme plastic behavior in terms of the tensile strength $\tensilestrength$, defined as the maximum stress reached by the stress-strain curve. The resulting values for the bulk and grain-boundary crystals are presented as a function of the atomic hydrogen concentration $\conc$ in Fig.~\ref{fig:modulus_and_yield_stress}(b). From there, we see that the addition of hydrogen decreases the tensile strength of both configurations. This can be understood by noting that the H atoms introduce disorder in the Fe lattice. As in the case of the Young's modulus, the decrease is more substantial in the bulk than in the grain-boundary system.  Without hydrogen, their tensile strengths are approximately equal ($\tensilestrength=7.31$\,GPa for the bulk and 7.27\,GPa for the grain-boundary system). At a hydrogen concentration of 9.1\%, however, the tensile strength of the bulk system has decreased by 54\%, while in the grain-boundary system the decrease is only 29\%. The reason for the weaker effect in the grain-boundary system is likely the same as mentioned above for the Young's modulus.

Introduction of hydrogen into the system also modifies the shape of the stress--strain curves near the maximum stress. For pure bulk iron, there is a sudden drop in the stress at 9\% strain~[Fig.~\ref{fig:stress--strain_bulk_gb}(a)]. Visual inspection~\cite{ovito} of the simulation system reveals that this is caused by a slip process that creates stacking-fault ribbons extending through the system in the $z$ direction. A similar but less distinct drop occurs for the grain-boundary system [Fig.~\ref{fig:stress--strain_bulk_gb}(b)]. In this case, the stacking fault cannot extend through the whole system because the grain boundary interrupts the crystal structure. The presence of hydrogen smooths the abrupt drops by significantly disturbing the crystal lattice already before the stacking-fault ribbons appear. 

\section{Discussion}\label{sc:discussion}

We have developed an analytical Tersoff--Brenner potential for interactions between hydrogen and iron atoms. It was fitted to a set of experimental and \emph{ab initio} data on iron hydride molecules, rock-salt-structured crystalline FeH, and hydrogen point defects in iron. The obtained potential reproduces the experimentally measured bond energy, bond length, and ground-state vibrational frequency of the FeH dimer and describes with good accuracy the molecules FeH${}_2$ and FeH${}_3$ as well as the rock-salt FeH. The point-defect energies it predicts are also consistent with our own DFT calculations.

The constructed potential enables atomistic computer simulations of a wide range of materials problems involving iron and hydrogen. Since it can also model nonequilibrium phenomena such as sputtering and the formation of mixed materials, the potential is well-suited for MD studies of plasma--wall interactions in fusion reactors. In view of the recent design updates of the ITER reactor, which would result in direct exposure of steel to the fusion plasma~\cite{CarbonoutfromITERICFRM}, being able to incorporate both iron and hydrogen into these investigations is an important advancement. With the Fe--H part now available, the only Tersoff--Brenner potential missing from the quaternary Fe--Cr--C--H system is Cr--H, whose development we leave for future work.

In Section~\ref{sc:tensile_test}, we applied the potential to tensile-test simulations of iron in two different configurations, a bulk bcc monocrystal and a symmetric tilt boundary, using different concentrations $\conc$ of hydrogen impurity atoms. The simulations indicated that hydrogen softens iron; i.e., the Young's modulus and the tensile strength decrease when the hydrogen concentration increases. The effect was much stronger in the bulk bcc monocrystal than in the tilt-boundary system. This was explained by noting that most of the hydrogen in the grain-boundary system was concentrated near the grain boundaries, thereby leaving the rest of the system depleted in hydrogen in comparison to the bulk system, where hydrogen was homogeneously distributed. 

Our simulations demonstrate that the potential can be used to study hydrogen-induced embrittlement phenomena in iron and steel. We emphasize that the current simulation setup is constructed as a simple model system for potential testing. In likely experimental scenarios, most H would (due to its low solubility) be trapped in defects or in grain boundaries. Thus, the potential's prediction of hydrogen-induced grain-boundary weakening is at least qualitatively consistent with the well-known effect of grain-boundary embrittlement by H in steels~\cite{Ber1970.MSE6.1,Zho2000.PRB62.13938,Nag2001.MetMatTransA32.339, Tat2003.PRB67.174105,Gen2005.MatTrans46.756,Tia2011.JPCM23.015501}. Future work could examine this more systematically for other grain boundaries and hydrogen distributions.

\section{Summary}\label{sc:summary}

We constructed a DFT-based interatomic potential for the Fe--H system in the Tersoff--Brenner formalism. The potential can be directly combined with our previously developed potential for the steel Fe--C system, to enable simulations of hydrocarbon chemistry in steel bulk and at steel surfaces. We applied the new potential to investigating the effect of hydrogen on the mechanical properties of monocrystalline bulk Fe and an Fe bicrystal with a grain boundary. In both cases, hydrogen was found to soften the material, reducing the Young's modulus as well as the tensile strength. 

\section*{Acknowledgments}
We thank T. Ahlgren, C. Bj\"orkas, F. Granberg, K. O. E. Henriksson, A. Lasa, and M. Nagel for insightful discussions. We are grateful for the computational resources granted by the CSC -- IT Center for Science in Espoo, Finland. P.~Kuopanportti acknowledges financial support from the Emil Aaltonen Foundation, the Finnish Cultural Foundation, the Magnus Ehrnrooth Foundation, and the Technology Industries of Finland Centennial Foundation. E.~Hayward and C.-C.~Fu thank the ANR-HSynThEx Project and GENCI (x2015096020). This project has been carried out within the framework of the EUROfusion Consortium and has received funding from the European Union's Horizon 2020 research and innovation programme under grant agreement number 633053. The views and opinions expressed herein do not necessarily reflect those of the European Commission.

\appendix

\section{Ab initio methods}\label{sc:appendix}
\setcounter{table}{0}

Here we outline our DFT calculations of the hydrogen point-defect energies in Table~\ref{table:defects}. They were performed with the {\footnotesize SIESTA} code~\cite{Sol2002.JPCM14.2745}, were spin-polarized within the collinear approximation, and used the generalized gradient approximation~(GGA) with the Perdew--Burke--Ernzerhof exchange-correlation functional~\cite{Per1996.PRL77.3865}. Core electrons were replaced by nonlocal norm-conserving pseudopotentials, and valence electrons were described by linear combinations of numerical pseudoatomic orbitals. We represented the charge density on a real-space grid with a spacing of $\sim\!\!0.07$\,\AA\ and employed a Methfessel--Paxton smearing~\cite{Met1989.PRB40.3616} of 0.3\,eV. All calculations used a 128-atom supercell with a $3\times 3\times 3$~$\mathbf{k}$-point grid. Zero-point energy corrections, calculated for hydrogen within the Einstein approximation, are included in our quoted values.

In Tables~\ref{table:properties} and~\ref{table:defects}, we also employed a number of abbreviations for the \emph{ab initio} methods used in other studies. The abbreviations are defined in Table~\ref{table:acronyms}.

\begin{table*}
\caption{\label{table:acronyms}Abbreviations used for the \emph{ab initio} methods in Tables~\ref{table:properties}
and~\ref{table:defects} (in alphabetical order).}
\begin{tabular}{ll}
\\                             
 Method\hspace{3pt} &Description \\ 
\hline \vspace{-5.0pt} \\ 
\BLYP &Becke's three-parameter hybrid functional~\cite{Bec1993.JCP98.5648} with\\
\vspace{4pt} &Lee--Yang--Parr correlation functional~\cite{Lee1988.PRB37.785}\\
\vspace{4pt}
\vspace{4pt}\CASSCF &Complete active space self-consistent field\\
\CCSDT &Coupled-cluster method in singles and doubles approximation\\
\vspace{4pt} &with connected triple excitation terms \\ 
\vspace{4pt}\CI &Configuration interaction \\ 
\vspace{4pt}\DFT &Density functional theory \\ 
\vspace{4pt}\ECP &Energy-adjusted quasi-relativistic effective core potential\\ 
\FLAPW &Full-potential linear-augmented-plane-wave method with \\
\vspace{4pt} &von~Barth--Hedin formula~\cite{Bar1972.JPC5.1629} \\ 
\vspace{4pt}\GGA &Generalized gradient approximation for DFT \\ 
\MBPPCA & Mixed-basis pseudopotentials with Perdew--Zunger\\ 
\vspace{4pt} &parametrization~\cite{Per1981.PRB23.5048} of Ceperley--Alder data~\cite{Cep1980.PRL45.566}\\
\vspace{4pt}\MRCPA &Fourth-order approximation of coupled pair approach \\ 
\MRSDCIQ &Single and double excitation multireference CI with Davidson\\ 
\vspace{4pt} & quadruple excitation correction \\ 
\vspace{4pt}\PAW &Projector-augmented-wave method within GGA \\ 
\SDQMBPT &Fourth-order many-body perturbation theory restricted to\\
\vspace{4pt} & single, double, and quadruple excitations \\ 
\vspace{4pt}\UHF &Unrestricted Hartree--Fock method\\ 
\vspace{4pt}\USPEX &Evolutionary crystal structure prediction method~\cite{Oga2006.JCP124.244704} (GGA) \\ 
\USPP &Ultrasoft pseudopotentials within GGA \\
\hline
\end{tabular}
\end{table*}

\bibliographystyle{apsrev4-1}
\bibliography{feh-manu-cms}

\begin{thebibliography}{86}
\bibitem [{not({\natexlab{a}})}]{note_extra_data}%
  \BibitemOpen
  \bibinfo {note} {{In
  Fig.~\ref{fig:modulus_and_yield_stress}, we also use data from the interval
  $0.05< n < 0.1$ (stress--strain curves not shown).}}\BibitemShut {Stop}%
\end{thebibliography}

\end{document}